\documentclass[
 reprint,
 amsmath,amssymb,
 aps,
]{revtex4-2}

\usepackage{graphicx}
\usepackage{dcolumn}
\usepackage{bm}
\usepackage{dsfont}

\begin{document}

\preprint{APS/123-QED}

\title{Geometric Control of Collective Spontaneous Emission}
\author{Yizun~He$^1$}
\email{yzhe16@fudan.edu.cn}
\author{Lingjing~Ji$^1$}
\author{Yuzhuo~Wang$^1$}
\author{Liyang~Qiu$^1$}
\author{Jian~Zhao$^1$}
\author{Yudi~Ma$^1$}
\author{Xing~Huang$^1$}
\author{Saijun~Wu$^1$}%
\email{saijunwu@fudan.edu.cn}
\affiliation{%
$^1$Department of Physics, State Key Laboratory of Surface Physics and Key 
Laboratory of Micro and Nano Photonic Structures (Ministry of Education), 
Fudan University, Shanghai 200433, China}
\author{Darrick~E.~Chang$^{2,3}$}
\email{darrick.chang@icfo.eu}
\affiliation{
$^2$ICFO-Institut de Ciencies Fotoniques, The Barcelona Institute of Science 
and Technology,
08860 Castelldefels, Barcelona, Spain and\\
$^3$ICREA-Instituci\'o Catalana de Recerca i Estudis Avan\c{c}ats, 08015 
Barcelona, Spain
}%


\date{\today}


\begin{abstract}
  Dipole spin-wave states of atomic ensembles with wave vector ${\bf k}(\omega)$ mismatched from the dispersion relation of light are difficult to access by far-field excitation but may support rich phenomena beyond the traditional phase-matched scenario in quantum optics. We propose and demonstrate an optical technique to efficiently access these states. In particular,  subnanosecond laser pulses shaped by a home-developed wideband modulation method are applied to shift the spin wave in ${\bf k}$ space with state-dependent geometric phase patterning, in an error-resilient fashion and on timescales much faster than spontaneous emission. We verify this control through the redirection, switch off, and recall of collectively enhanced emission from a $^{87}$Rb gas with $\sim 75\%$ single-step efficiency.  Our work represents a first step toward efficient control of electric dipole spin waves for studying many-body dissipative dynamics of excited gases, as well as for numerous quantum optical applications.
\end{abstract}
\maketitle



Since the seminal work by Dicke in 1954, it has been well known that collective effects can significantly modify light emission from an ensemble of excited atoms~\cite{Dicke1954}. In the case of the original Dicke model, it was predicted that colocated atoms could experience super- or subradiance, where the emission rate is enhanced or even completely suppressed depending on the correlations of the excited atoms. Today, collective enhancement forms the cornerstone for applications based upon atomic ensembles, such as quantum memories~\cite{Lukin2000, Fleischhauer2005} and quantum repeater protocols~\cite{DLCZ2001} for single-photon generation. An important concept underlying these ideas is that a shared, collective atomic excitation, in the form of an electric dipole spin wave with well-defined wave vector ${\bf k}$, will radiate efficiently into a narrow spatial mode centered around the direction of ${\bf k}$ rather than into all directions, when the wave vector is phase matched to the dispersion relation of electromagnetic radiation with $|{\bf k}|=\omega/c$ ~\cite{scully2006,Eberly2006}. 

More recently, there has been growing interest in dipole spin-wave states with $|{\bf k}| \neq \omega/c$ strongly mismatched from radiation and thus suppressed collective emission, which support completely new opportunities within quantum optics. For example, the physics emerging from dipole-dipole interactions and multiple scattering in dense atomic gases is a complex question of active interest~\cite{Morice1995,Bons2016,
Skipetrov2014, Bellando2014, Jennewein2016, Schilder2016,  Corman2017,  oeJavanainen2017,  Jennewein2018,  Skipetrov2018,  Schilder2020,  RGChang2020}. There is increasing evidence that the subradiant modes of nearby atomic pairs play an important role in some of these phenomena, and large wave vector spin waves could be used to efficiently probe and manipulate these modes. Furthermore, 
phase-mismatched spin waves have been predicted to have highly subradiant character in ordered atomic arrays where interference in light emission is strongly enhanced. Such states could lead to intriguing linear optical or many-body phenomena~\cite{  Ritch2010,  Du2015,  Adams2016,  Bettles2016,  syznanov2016, Perczel2017,  Shahmoon2017,  Bettles2017,  Molmer2019,  Needham2019,Bettles2019}, or be an important resource in applications~\cite{Ritch2013, Ritch2016,    selectiveRadiancePRXchang17,    Guimond2019,prl2016,  DEC2019PRA}. 

Of course, techniques to manipulate spin waves based upon Raman transitions and multiple long-lived ground states are ubiquitously used within atomic physics~\cite{Lukin2000, Fleischhauer2005,DLCZ2001,SIMON12007,NPZhao2009,Xiao2012,Duan2019}, and in principle could be adapted to indirectly generate mismatched dipole spin waves on optical transitions. However, in some cases it would be highly desirable to generate such excitations by working directly on a strong transition without the need for additional ground states. First, direct electric dipole control ensures the fastest possible manipulation, with benefits such as to achieve high efficiency with suppressed low-frequency noise. Furthermore, many subradiant effects as in Refs.\cite{Ritch2010,  Du2015,  Adams2016,  Bettles2016,  syznanov2016,
Perczel2017,  Shahmoon2017,  Bettles2017,
Molmer2019,  Needham2019, Bettles2019, Ritch2013, Ritch2016,    selectiveRadiancePRXchang17,    Guimond2019,prl2016,  DEC2019PRA} are only known to occur in ``closed'' transitions with unique ground states, and possible generalization to multiple ground states is still largely unknown~\cite{Hebenstreit17, AsenjoGarcia19}. A number of techniques have been proposed to directly shift the spin-wave vector in the time domain, on timescales much faster than the dipole decay itself. These include using dynamic or geometric phases with ultrafast pulses~\cite{scully2015}, applying Zeeman shifts or vector light shifts~\cite{prl2016}, or generating state-dependent spatially periodic dynamic phases~\cite{Ballantine2020}. To date, however, achieving the high efficiencies needed to open up new opportunities remains an outstanding experimental challenge.

\begin{figure*}
\includegraphics[width=1.0\textwidth]{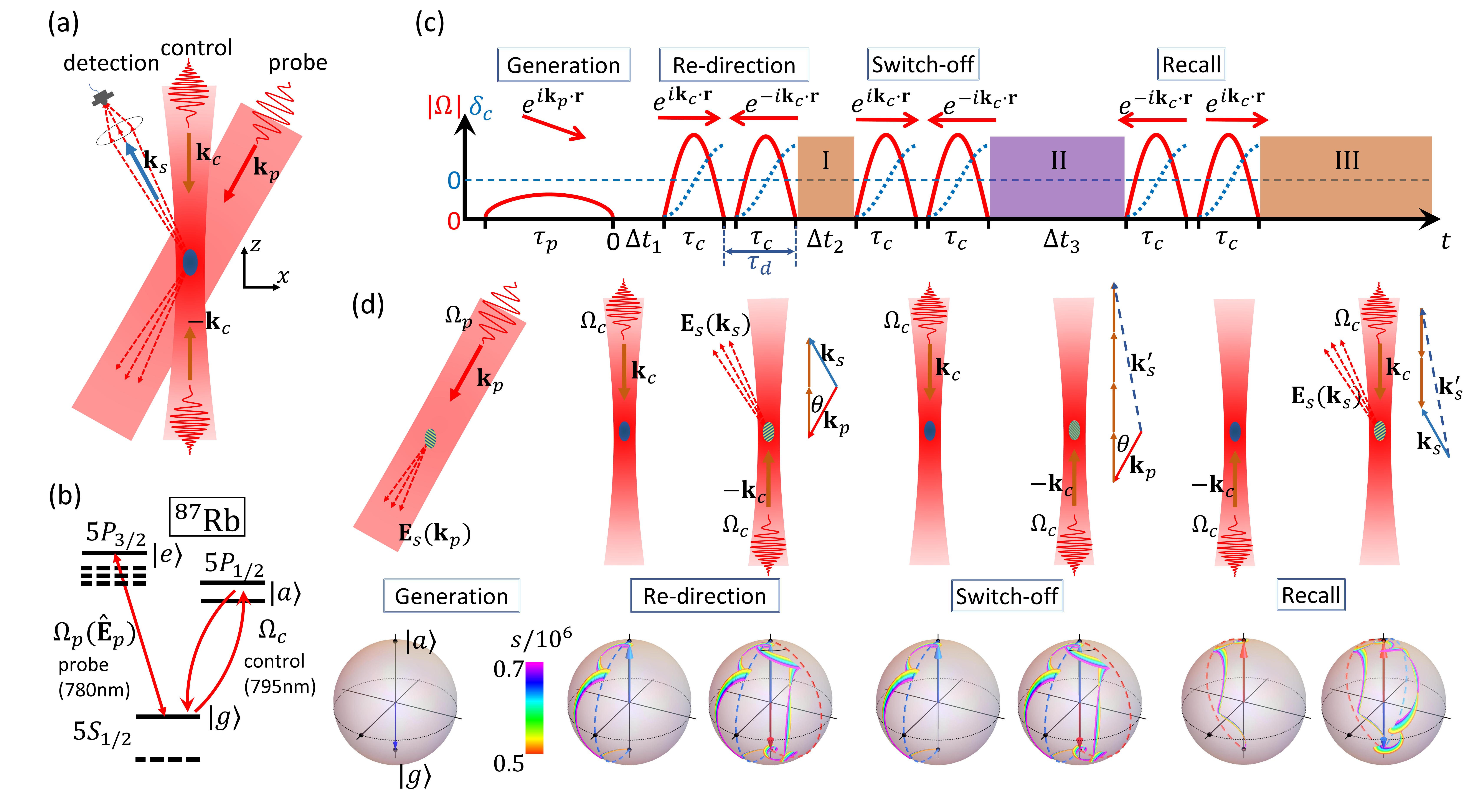}
\caption{Schematic of the experiment to demonstrate error-resilient dipole 
spin-wave control. (a) The basic setup. (b) The level diagram and the laser 
coupling scheme. (c) Schematic timing sequence for the amplitudes of the 
probe and control Rabi frequency $|\Omega|$ (red solid lines), and the 
instantaneous detuning of the control pulse $\delta_c$ (blue dashed lines) 
from the $|g\rangle -|a\rangle$ transition. (d) Top (from left to right): 
generation and control of dipole spin wave and the redirection, switch off and recall of the collective spontaneous emission. The 
$|g\rangle-|e\rangle$ electric dipole spin wave is illustrated with fringes 
in the sample. The drawings are not to actual scale, in 
particular, the phase-matching angle 
$\theta=\arccos\frac{|{\bf k}_c|}{|{\bf k}_p|}\sim 11.1^{\circ}$ is exaggerated for 
clarity. Bottom: Bloch-sphere 
representation of the projected $|g\rangle-|a\rangle$ dynamics for an
atom at position $\bf r$. Calculated trajectories with different control pulse 
peak intensity parameter $s$ are displayed. The quasiadiabatic control 
ensures the intensity error resilience: The geometric phase writing is insensitive to small deviations of $s$ from
$s\sim 0.6\times 10^6$, for values of $\tau_c \Gamma_{D1}$ between 0.02 and 0.03 in this work.}\label{figSetup}
\end{figure*}

In this Letter, we demonstrate a general method to coherently shift the wave vectors of dipolar spin waves on a strong transition to achieve efficient conversion between phase-matched and -mismatched states. 
The rapid ${\bf k}$-space shift is made possible by a home-developed technique to generate subnanosecond shaped laser pulses that we use to cyclically drive~\cite{accMetcalf2007} an auxiliary transition~\cite{scully2015}, on a timescale much faster than the dipole decay. Importantly, this subwavelength geometric phase patterning technique is robust to imperfections in detuning and intensity variations across the beam, which is essential for achieving precise spin-wave control with a focused laser. We benchmark this control by finely adjusting the ${\bf k}$-space shift to achieve a redirection of collective emission from a phase-matched spin wave, followed by a coherent shift to $|{\bf k}|>\omega/c$ off the light cone with single-operation efficiency at the $75\%$ level, and finally a shift back to a phase-matched spin wave on demand. The present control efficiency is limited by atomic hyperfine interactions and spontaneous decay~\cite{coSub}, and we expect substantial improvements by equipping an optical pulse shaper with higher bandwidth than the one developed for this work~\cite{Yudi2020}.
Our method should lead to multiple new opportunities within atomic physics and quantum optics, such as strong coherent atomic acceleration for atom interferometry~\cite{Bouchendira2011,freegarde2014,Stoner2015,
Mueller2018,DaWei2014superAI}, and to unlock rich physics associated with collective phenomena in disordered and ordered atomic systems, both in the linear optical~\cite{Skipetrov2014, Bellando2014, Jennewein2016, Schilder2016,  Corman2017,  oeJavanainen2017,  Jennewein2018,  Skipetrov2018,  Schilder2020,  RGChang2020, Ritch2010,  Du2015,  Adams2016,  Bettles2016,  syznanov2016, Perczel2017,  Shahmoon2017,  Bettles2017,Needham2019,Guimond2019,prl2016,selectiveRadiancePRXchang17}
and quantum many-body~\cite{Ritch2013, Molmer2019, Bettles2019,DEC2019PRA,DarrickNP2014Rev,ChangsukRPP2017Rev} regimes. We also note that high-fidelity control of spin ensembles over the Bloch sphere constitutes the cornerstone of nuclear magnetic resonance (NMR)~\cite{NMRAdiabatic2001, NatCommDu2015,Chuang2016}, and our work potentially brings ideas developed in that field to bear on the completely new regime of fast-decaying atomic transitions.




As depicted in Fig.~\ref{figSetup}, we work with laser-cooled $^{87}$Rb atoms on the $D1$ and $D2$ strong optical transitions, with $|g\rangle$,$|e\rangle$ and auxiliary $|a\rangle$ states corresponding to the 5$S_{1/2}~F=2$, 5$P_{3/2}~F'=3$, and 5$P_{1/2}~F'=1,2$ hyperfine levels, respectively. The nondegenerate gas contains $N\sim 5\times10^3$ atoms at $T=20~\mu$K over an $\sim15~\mu$m spatial region (peak density $\sim 4\times 10^{12}/{\rm cm}^3$)~\cite{supplementalmaterial,coSub}. Natural linewidths of the $D1$ and $D2$ transitions are $\Gamma_{D1} = 2\pi \times 5.75$~MHz and $\Gamma_{D2} = 2\pi \times 6.07$~MHz, respectively. A spatially uniform $D2$ spin-wave excitation (or more precisely, a weak coherent state involving such excitations) is generated by an incoming plane-wave ``probe pulse'' with frequency resonant to the transition frequency $\omega_{eg}$ and duration $\tau_p$ short enough that light rescattering effects are negligible during the interaction. The spin-wave states are described by ``timed-Dicke'' states of the form $|\psi_{\rm TD}({\bf k}_p)\rangle=S^+({\bf k}_p)|g_1,g_2,...,g_N\rangle$. Here, $S^+({\bf k}_p)=\frac{1}{\sqrt{N}}\sum_i{e^{i {\bf k}_p\cdot {\bf r}_i} 
|e_i\rangle\langle g_i|}$ denotes a collective spin-raising operator and ${\bf k}_p$ the probe beam wave vector. For a macroscopic ensemble it is well known that the ${S^+}({\bf k}_p)$ excitation with $|{\bf k}_p|=\omega_{eg}/c$ emits light along the phase-matched ${\bf k}_p$ direction in a superradiant fashion~\cite{scully2006,Eberly2006}, with collectively enhanced emission rate and shifted optical frequency.
For the dilute gas in this work, the enhancement leads to $N$-fold increase of forward emission power, and a collective damping of excitation at a rate proportional to the resonant optical depth along ${\bf k}_p$~\cite{superradianceAraujo16,superradianceRoof2016,coSub}. However, in this work, we are interested in spin waves with strongly mismatched $|{\bf k}|\neq \omega_{eg}/c$, in which case the collective emission is prohibited. In an ensemble with random atomic positions $\{{\bf r}_i\}$, as in this work, such a spin wave is expected to decay with a rate near the natural linewidth~\cite{scully2015}. On the other hand, in an ordered array of atoms, the destructive interference in all directions can be nearly perfect, leading to strong subradiance~\cite{Ritch2010,  Du2015,  Adams2016,  Bettles2016,  syznanov2016,
Perczel2017,  Shahmoon2017,  Bettles2017,
Molmer2019,  Needham2019, Bettles2019, Ritch2013, Ritch2016,    selectiveRadiancePRXchang17,    Guimond2019,prl2016,  DEC2019PRA}.

\begin{figure}
    \includegraphics[width=0.4\textwidth]{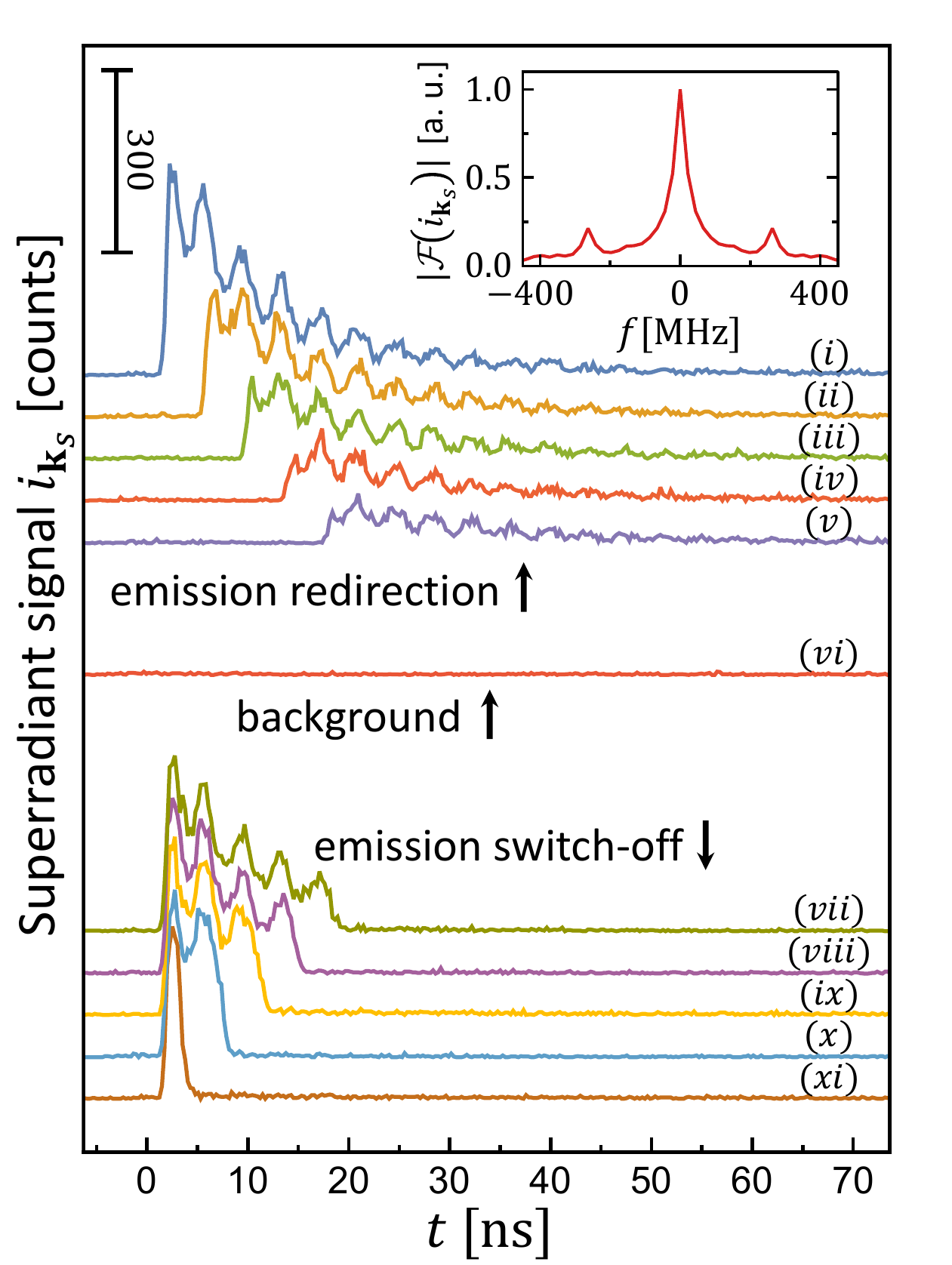}
    \caption{Typical signals of superradiant emission $i_{{\bf k}_s}$ under coherent dipole spin-wave control. The probe pulse excites the system in the interval $-\tau_p<t<0$, with $\tau_p=3$~ns. Curves (i)-(v) show the signal collected along the ${\bf k}_s$ direction, which exhibit a steep rise once the initial spin wave with wave vector ${\bf k}_p$ is redirected along this direction at the delay times $\Delta t_1$=0.0, 4.0, 8.0, 12.0, 16.0~ns. In curves 
    (vii)-(xi), we redirect the superradiance at $\Delta t_1= 0.0$~ns, and subsequently perform a switch off at delay times $\Delta t_2=$16.6, 12.6, 8.6, 4.6, 0.6~ns, where the spin wave is shifted to a phase-mismatched amplitude of $|{\bf k}_s'|\approx 2.9 \omega_{eg}/c$. It can be seen that the emission signal strongly shuts off at the expected times $\Delta t_2$. The top right inset gives 
    the Fourier transform of curve (i). In curve (vi), the $D2$ excitation is implemented without any spin-wave control, with a tiny amount of photons from scattering of the probe beam and emission of atoms detected. By avoiding nearby optics and through optical filtering, the control light background is completely suppressed from the signals.}\label{figSuper}
\end{figure}

To access the phase-mismatched spin waves, we send a pair of counterpropagating chirped $D1$ pulses with subnanosecond duration $\tau_c$ and relative delay $\tau_d$ to the sample, immediately after the $D2$ $S^+({\bf k}_p)$ spin-wave preparation [see Fig.~\ref{figSetup}(c)].
For the $D1$ control pulse, the peak saturation parameter is $s \sim 10^6$ ($s\equiv$ intensity$/I_{s1}$ and $I_{s1}= 4.49~$mW/cm$^2$ is the $D1$ transition saturation intensity), with the peak Rabi frequency being $\Omega_0=\sqrt{s/2}\Gamma_{D1}$ which is at the GHz level.
The waveform-optimized pulse pair with  opposite wave vector $\pm{\bf k}_c$ drives the auxiliary $|g\rangle\rightarrow |a\rangle$ and $|a\rangle\rightarrow |g\rangle$ transitions successively via two rapid adiabatic passages~\cite{accMetcalf2007} separated by the delay $\tau_d$. Meanwhile, the state $|e\rangle$ is affected negligibly by the control pulses due to the large $D1$-$D2$ optical frequency difference. As detailed in the Supplemental Material~\cite{supplementalmaterial}, such a 2$\pi$ rotation imparts a geometric phase $\varphi_{G}=\pi\pm 2{\bf k}_c\cdot {\bf r}_i$ on each atom in $|g\rangle$ given by half the solid angle subtended on the Bloch sphere, which is different for each atom due to the local phase factors $e^{\pm i {\bf k}_c \cdot {\bf r}_i}$ seen by each atom in the two control pulses. It can readily be shown that the position and ``spin''-dependent phase acquired is exactly equivalent to a shift of the spin-wave excitation, $S^{+}({\bf k}_p) \rightarrow S^{+}({\bf k}_p\mp 2{\bf k}_c)$ if the $\pm {\bf k}_c$ control pulse arrives first to drive the $|g\rangle\rightarrow |a\rangle$ transition. Starting from the phase-matched $S^+({\bf k}_p)$ excitation, multiple ${\bf k}\rightarrow {\bf k}\pm 2{\bf k}_c$ shifts generally lead to mismatched spin waves with $|{\bf k}_p\pm 2 n{\bf k}_c|\neq\omega_{eg}/c$.  

We have optimized the control pulses so that the phase imprinting is robust to intensity across the laser beam, using chirped control waveforms [Fig.~\ref{figSetup}(c)] with Rabi frequency $\Omega_c(t)=\Omega_0 \sin(\pi t/\tau_c)$ and instantaneous detuning $\delta_c(t)=-\delta_0 \cos(\pi t/\tau_c)$ (defined relative to the midpoint of the 5$S_{1/2}~F=2$ -- 5$P_{1/2}~F'=1,2$ hyperfine lines). The optimization is carried out experimentally and confirmed numerically~\cite{coSub}. While such geometrically robust control techniques are well known in, e.g., nuclear magnetic resonance~\cite{NMRAdiabatic2001}, an obvious challenge when applied to atomic optical transitions is the rapid decay time of the atomic excited state, which requires one to precisely shape subnanosecond pulses of sufficiently high power. It is worth pointing out that to isolate the two-level optical spin from uncontrolled multilevel couplings, the shaped pulses should be nearly resonant and not too fast~\cite{GoswamiPR2003,coSub,Yudi2020,campbell2019,Heinrich2019}. As detailed in the Supplemental Material~\cite{supplementalmaterial}, we have developed a new instrumentation based on high-speed sideband modulation of an amplified cw laser to meet the requirements for the flexibly programmable error-resilient control.


Our goal is to demonstrate coherent conversion between matched and mismatched spin waves, with only the former being efficiently detectable by light emission. Furthermore, for a weakly excited sample and moderate total atom number $N$, the detection needs to be efficient at the single-photon level. While the initial spin wave along ${\bf k}_p$ generated by the probe pulse is already phase matched, detection of the light emitted along ${\bf k}_p$ is problematic as the detector would be blinded by the strong probe pulse. We thus first take advantage of our technique to perform a redirection after a delay time $\Delta t_1$ following the generation of the initial spin excitation $S^{+}({\bf k}_p)$. In this redirection, we convert this excitation to another $S^{+}({\bf k}_s={\bf k}_p-2{\bf k}_c)$, which is also phase matched, $|{\bf k}_s|=\omega_{eg}/c$. This is made possible by finely aligning the angle between ${\bf k}_p$ and ${\bf k}_c$ (Fig.~\ref{figSetup}), and the photon collection optics are correspondingly aligned along the ${\bf k}_s$ direction. We then perform a switch off of emission, where after a delay time $\Delta t_2$, we perform a shift to realize a highly mismatched spin wave $S^{+}({\bf k}_s'={\bf k}_s-2{\bf k}_c)$, with $|{\bf k}_s'|\approx 2.9 \omega_{eg}/c$. Finally, we realize a recall by performing a shift back to $S^{+}({\bf k}_s)$ after a delay time $\Delta t_3$. Notably, the emission we observe in this final state also reflects any dynamics accumulated by the mismatched spin wave during the time $\Delta t_3$. This would be particularly valuable, e.g., in an atomic array, where the dynamics is predicted to be extremely rich.

\begin{figure}
    \includegraphics[width=0.46\textwidth]{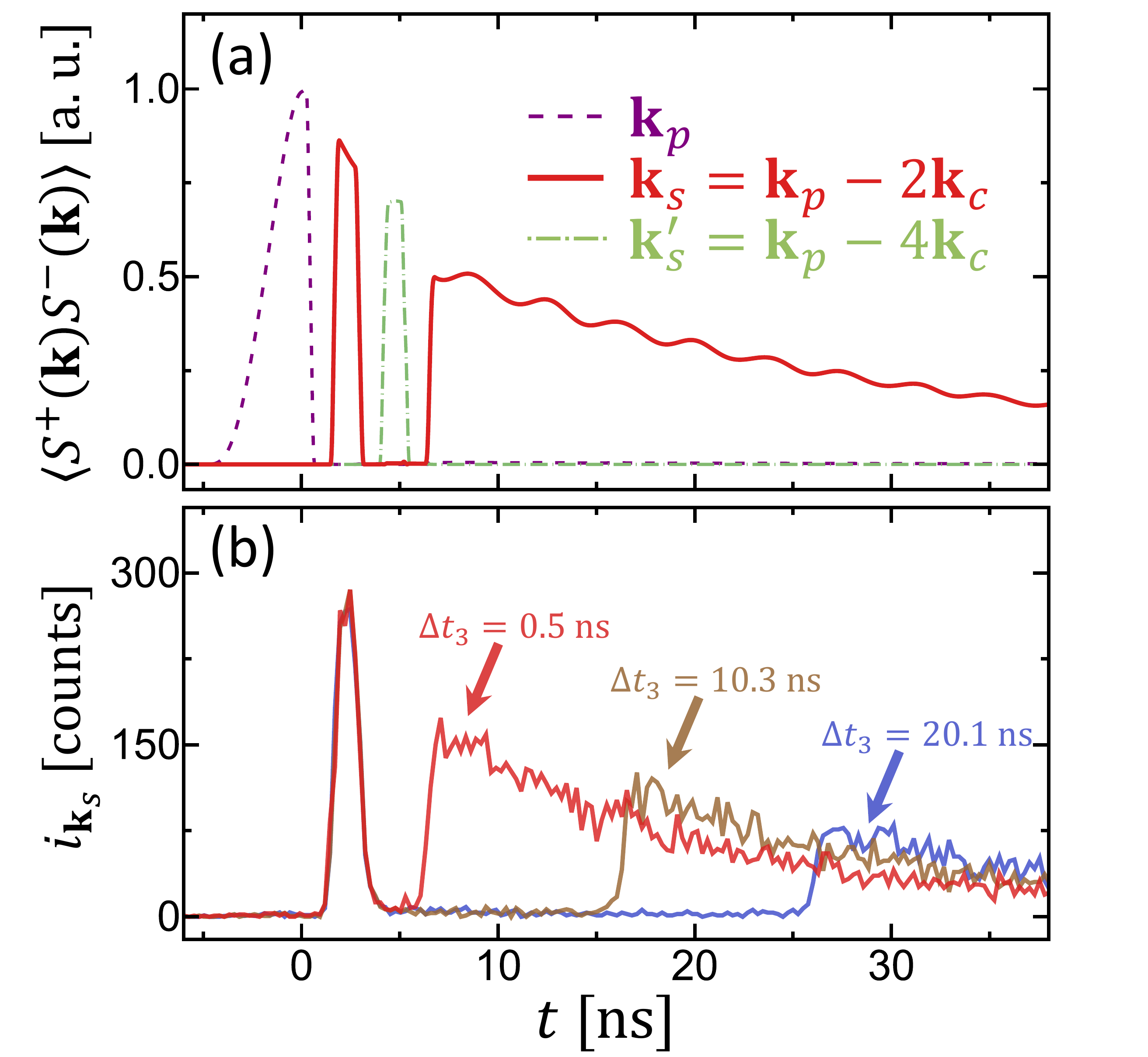}
    \caption{Controlled spin-wave dynamics and collective spontaneous emission $i_{{\bf k}_s}$ by the $S^+({\bf k}_s)$ excitation during a redirection--switch-off--recall sequence. 
    (a) Simulation of the spin-wave population $\langle S^+({\bf k}_p) S^-({\bf k}_p)\rangle$ with wave vector corresponding to the probe beam (purple dashed line), the intensity of light $i_{{\bf k}_s}\propto \langle S^+({\bf k}_s) S^-({\bf k}_s)\rangle$ emitted in the direction along the redirected wave vector ${\bf k}_s$ (red line), and the mismatched spin-wave population $\langle S^+({\bf k}_s') S^-({\bf k}_s')\rangle$ (green dot-dashed line). These simulations are based on the optical Bloch equations with estimated experimental parameters detailed in the Supplemental Material~\cite{supplementalmaterial}. (b) Experimentally measured photon emission counts $i_{{\bf k}_s}$ with different recall delay $\Delta t_3=0.5, 10.3$, $20.1$~ns. $\Delta t_1 = 0.2$~ns and $\Delta t_2 = 0.8$~ns are fixed. The control pulse parameters are $\tau_c=0.5$~ns, $\tau_d=1.2$~ns, $\Omega_0 \approx 2\pi\times 4$~GHz and $\delta_0 \approx 2\pi\times 4$~GHz. With a longer duration $\tau_p = 5$~ns for the probe excitation, the quantum beats as those in Fig.~\ref{figSuper} are suppressed.
    \label{figsample} }
\end{figure}




The dipole spin-wave control is first demonstrated by the emission redirection as in curves (i)-(v) of Fig.~\ref{figSuper}, for various delay times $\Delta t_1$ of the application of the control pulses. Collective emission along the new ${\bf k}_s$ direction leads to a rapid rise of the photon counting curves at the expected $\Delta t_1$. The $\sim 1$~ns rise time of the curves reflects the time needed for the second pulse in the control pair to bring the atoms in state $|a\rangle$ back to $|g\rangle$, imparting the desired geometric phase. 
The control parameters here are chosen to be $\tau_c=0.9$~ns, $\tau_d=1.36$~ns, $\Omega_0 = 2\pi\times 3$~GHz, and $\delta_0=2\pi\times 3.4$~GHz. The rise time is slightly broadened from $\tau_c=0.9~$ns due to counter timing jitters. The $f\approx 267$~MHz oscillations observed in the curves are due to a quantum beat between the $D2$ collective spontaneous emission from the $F'=3$ and those from the off-resonantly excited $F'=2$ levels~\cite{CsSpecWieman1994}. The superradiant emission signals decay on a timescale of $\tau_N\approx 12.3$~ns, revealing a collectively enhanced decay rate in forward emission~\cite{scully2006,Eberly2006,superradianceAraujo16,superradianceRoof2016}  (compared to a single-atom decay time of $\Gamma_{D2}^{-1}=26.2$~ns). The dependence of this decay time on the optical depth is detailed in Ref.~\cite{coSub}. The divergence angle of the ${\bf k}_s$-mode emission is estimated in the far field to be at the 20~mrad level, which is consistent with the diffraction limited angle of $\theta \sim 1/|{\bf k}_p|\sigma$ ~\cite{Gross1982} of our nearly spherical atomic sample with $\sigma\sim 7~\mu$m radius.

We proceed further with our spin-wave control by applying a second pair of switch-off control pulses for an additional ${\bf k}\rightarrow {\bf k}-2{\bf k}_c$ shift at various delay times $\Delta t_2$. With the spin-wave vector shifted to ${\bf k}_s' = {\bf k}_p-4{\bf k}_c$ to be strongly mismatched from radiation, the collective emission along ${\bf k}_s$ is effectively extinguished. Indeed, the random emission into the same $\theta$ detection angle by the mismatched spin-wave excitation, without the collective enhancement, is decreased by a factor of $\sim N$~\cite{scully2015,coSub}. The switch-off effect is clearly seen by the disappearance of the signal in curves (vii)-(xi) of Fig.~\ref{figSuper} at the associated $\Delta t_2$, with the $\sim 1$~ns falling edges of the curves reflecting the time over which the first pulse in the second control pair drives atomic population out of state $|g\rangle$.   

To fully verify the coherent nature of the spin-wave control, we finally perform a recall operation with a ${\bf k}\rightarrow {\bf k}+2{\bf k}_c$ shift to bring back the mismatched ${S^+}({\bf k}_s')$ spin wave to ${\bf k}_s$, allowing the atomic ensemble to emit collectively again. The effect is clearly demonstrated in Fig.~\ref{figsample}(b) where the photon counting curves record the full history of ${S^+}({\bf k}_s)$ emission during the redirection--switch-off--recall sequence. In particular, in each of the curves the first peak corresponds to the superradiant interval ``I'' in Fig.~\ref{figSetup}(c), with duration $\Delta t_2=0.8$~ns in these measurements, before the emission is switched off and the mismatched spin wave evolves in the dark [the interval ``II'' of Fig.~\ref{figSetup}(c)]. The second peak in intensity corresponds to the recalled emission after the $\Delta t_3$ delay. The collectively enhanced decay lifetime of the recalled emission signal is $\tau_N\approx 14.6$~ns for this specific sample. On the other hand, the decay of mismatched spin wave in the dark is close to the single-atom $\Gamma_{D2}$ rate~\cite{scully2015,coSub}, as can be inferred from the maximum amplitudes of the recalled signals for different delay times $\Delta t_3$. The spin-wave control thus allows us to generate the mismatched spin waves and to map them back to light on demand for the first time to our knowledge. 


We have simulated spin-wave dynamics supported by noninteracting atoms, including all the hyperfine structures of the $D1$ and $D2$ transitions, in order to better understand the experiment~\cite{supplementalmaterial}. This approach is valid in the limit that the resonant dipole interactions between atoms are negligible during the nanosecond spin-wave control, as we expect here. As illustrated in Fig.~\ref{figsample}(a), the simulation reproduces well the key features of the experimental observable $i_{{\bf k}_s}(t)\propto \langle S^+({\bf k}_s) S^-({\bf k}_s)\rangle$, and also unveils time-dependent dynamics for the unmonitored forward emission $\langle S^+({\bf k}_p) S^-({\bf k}_p)\rangle$  and the superradiance-free excitation 
$\langle S^+({\bf k}_s') S^-({\bf k}_s')\rangle$. To estimate the spin-wave control efficiency, we fit the amplitude of the recalled emission and compare that with the amplitude of the redirected emission in the same sequence as in Fig.~\ref{figsample}(b), which suggests a storage-recall efficiency for the controlled dipole spin-wave energy of $\sim 58\%$. The efficiency for a single ${\bf k}\rightarrow {\bf k}\pm 2{\bf k}_c$ shift is thus at the $75\%$ level, which is also consistent with a separate, systematic analysis of measurements of the optical acceleration induced by the spin-wave conversion~\cite{coSub}.

As analyzed in Ref.~\cite{coSub}, the fidelity of the spin-wave control in this work is primarily limited by the $m_F$-dependent hyperfine phase shifts for atoms initially in different Zeeman sublevels, and in addition by both $D1$ and $D2$ spontaneous emission during the $\tau_c+\tau_d \sim 2$~ns control time. With better initial atomic state preparation, such as optically pumping all the $^{87}$Rb atoms into the $F=2, m_F=0$ ground state, numerical simulations suggest we should expect $f_d \sim 87\%$ dipole spin-wave control efficiency. Instead of imparting the geometric phase to the ground state atoms, in future work an $|e\rangle-|a\rangle$ transition with a longer $|a\rangle$ lifetime~\cite{multiphotonMOT} may be chosen to implement $|e\rangle$-state phase patterning. The influence from the $D1$ decay can thus be eliminated, leading to $f_d\sim 95 \%$ limited by the suppressed $D2$ decay. With an additional fivefold reduction of $\tau_c+\tau_d$ to $400~$ps, we would expect $f_d$ reaching $99\%$ for high-fidelity dipole spin-wave control in dilute gases~\cite{coSub}. More generally, the imperfections of the control stemming from single-atom effects are manageable with better quantum control techniques~\cite{StAChen2010,freegarde2014,NatCommDu2015,Chuang2016} well developed in other fields, if they can be implemented in the optical domain with a reliable pulse shaping system of sufficient precision, bandwidth, and power~\cite{Yudi2020,Weiner2011Review}.



We expect only moderate reduction of control fidelity when our technique is applied to samples with higher densities, such as in the subwavelength array scenarios~\cite{Bettles2016, Shahmoon2017,Manzoni2018} as recently demonstrated in Ref.~\cite{Immanuel2020}. We suggest the imparted phase error $\delta\varphi$ due to atom-atom interactions in a dense atomic sample is bounded by $\delta \varphi_{\rm max}\sim \sqrt{\delta_N^2+\Gamma_N^2/4} \tau_c$~\cite{supplementalmaterial}, with $\delta_N$, $\Gamma_N$ being the largest collective Lamb 
shift~\cite{clScullyPRL09,Kaiser2016} and collective decay rate~\cite{scully2006} for the single spin-wave excitation under consideration. 


The spin- and position-dependent phase  $\sim e^{2 i {\bf k}_c \cdot {\bf r}}$ imparted in the control also results in a spin-dependent momentum kick of $\sim 2 \hbar {\bf k}_c$. The coherent acceleration effect is another topic to be addressed in full detail in Ref.~\cite{coSub} and may find applications at the interface of quantum optics and atom interferometry~\cite{Bouchendira2011,freegarde2014,Stoner2015,
Mueller2018,DaWei2014superAI}. For a free gas as in this work, the state-dependent acceleration introduces Doppler phase broadening to the spin waves, thus limiting the coherence time of the collective excitation to  $\tau_D\sim l_c/v_k$ where $l_c$ is the thermal de Broglie wavelength and $v_k=\hbar k/m $ is the recoil velocity associated with the $S^+({\bf k})$ collective excitation. In this work, with $l_c\approx 100$~nm for $^{87}$Rb at 20~$\mu$K and $v_k\approx 6$~mm/s, the Doppler dephasing time $\tau_D\sim 15~\mu$s does not affect the spin-wave dynamics at the $1/\Gamma_{D2}$ time scale. However, to maintain the phase coherence of the long-lived subradiant excitation in future work, particularly for lighter atoms with larger $v_k$, or for narrower line transitions with smaller natural linewidth~\cite{motionEffectsBromley2016}, the atoms should be confined by optical lattices in the recoilless or Lamb-Dicke regime, with the optical lattices at a ``magic'' wavelength and nullified dipole transition frequency shifts.



We note that time-domain control of coherent emission is also realized in ultrafast spectroscopy~\cite{cundiff2013,Oliver2018} for perturbative measurements of nonlinear polarizabilities. In contrast, here an efficient and precise control of collective spontaneous emission is realized, for the first time to our knowledge, by cyclically driving another transition so as to shift the phase-matching condition of the radiating dipoles on demand. The  geometric robustness is especially important for perfecting the optical spin control in macroscopic samples~\cite{GoswamiPR2003, coSub,Yudi2020} toward NMR-type precision~\cite{NMRAdiabatic2001, NatCommDu2015,Chuang2016} with a focused laser. With the outlined improvements~\cite{coSub}, we hope the nearly perfect dipole spin-wave control will find broad impact in atomic physics and quatum optics. Foremost, we envision the technique can become an important tool in investigating collective linear optical and many-body behavior of interacting dipoles beyond the traditional phase-matching scenario, both in disordered ensembles and atomic arrays. In the case of many-body behavior, for example, our technique would provide a means for multiexcitation of subradiant states. Under certain conditions, rich phenomena such as the buildup of ``fermionic'' correlations might arise in dynamics~\cite{Molmer2019,DEC2019PRA}, which can then be mapped back to light with high efficiency by a subsequent application of our technique. Such processes could also constitute a novel approach to realizing quantum nonlinear optics and the generation of exotic quantum many-body states of light~\cite{DarrickNP2014Rev,ChangsukRPP2017Rev}.

\begin{acknowledgments}
We are grateful to Professor Lei Zhou for both helpful
discussions and for kind support and to Professor J. V. Porto
and Professor Da-Wei Wang for helpful discussions and
insightful comments on the manuscript. We thank Professor
Kai-Feng Zhao and Professor Zheng-Hua An for help on
developing pulse shaping technology and for support from
the Fudan Physics nanofabrication center. D. E. C. acknowledges support from the European Union’s Horizon 2020
research and innovation program, under European Research
Council Grant Agreement No. 639643 (FOQAL) and Grant
Agreement No. 899275 (DAALI), MINECO Severo Ochoa
Program No. CEX2019-000910-S, Generalitat de Catalunya
through the CERCA program and QuantumCat (Reference
No. 001-P-001644), Fundacio Privada Cellex, Fundacio
Mir-Puig, Fundacion Ramon Areces Project CODEC, and
Plan Nacional Grant ALIQS funded by Ministerio de
Ciencia, Innovaci\'on y Universidades, Agencia Estatal de
Investigaci\'on, and European Regional Development Fund.
This research is mainly supported by National Key Research
Program of China under Grants No. 2016YFA0302000
and No. 2017YFA0304204, by NSFC under Grant
No. 11574053, and by Shanghai Scientific Research
Program under Grant No. 15ZR1403200.
\end{acknowledgments}

\appendix
\renewcommand*{\thesection}{\Alph{section}}

\section{Experimental details\label{AppA}}

\subsection{High speed pulse shaping system}
\begin{figure} 
\includegraphics[width=0.47\textwidth]{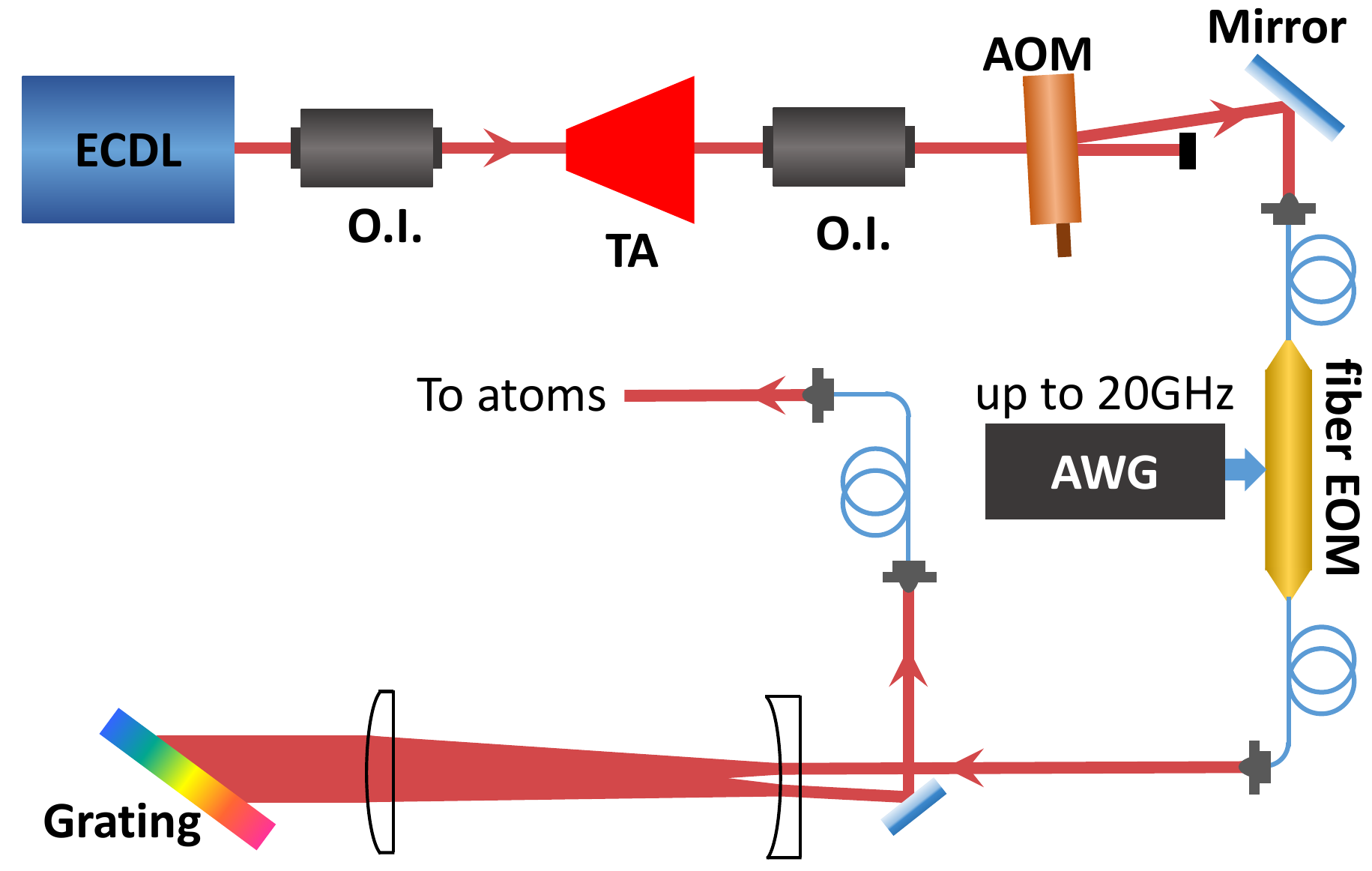} 
\caption{Schematic layout of the sub-nanosecond OAWG. An external cavity 
diode laser (ECDL) seeds a tapered amplifier (TA) which is followed by an 
acousto-optical modulator~(AOM) to pulse the input of the fiber 
electro-optical phase modulator~(fEOM). The fEOM is driven by a microwave 
arbitrary waveform generator~(AWG). The fEOM output is frequency-filtered by 
a large-area grating, by coupling the -1st order grating diffraction to a 
single-mode fiber. The output of the polarization-maintaining fiber is 
focused to the cold atoms samples with a beam waist of $w\approx 13~\mu$m as 
control light, which is shown in Fig.~1(a) of the main text. The O.I. is for 
optical isolators.} \label{figlasersetup}
\end{figure}

We develop a high speed pulse shaping system to generate both the 
sub-nanosecond $D1$ control pulse and nanosecond $D2$ probe pulse in this work. 
The system is based on fiber-based electro-optical modulation of an 
optical amplified external cavity diode laser (ECDL) 
output~\cite{Gould2016}, followed by a grating based optical filter. A 
simplified schematic setup is given in Fig.~\ref{figlasersetup}.

For the $D1$ control pulse generation, the ECDL is offset-locked to the 
hyperfine crossover between the $F=2- F'=1$ and $F=2 - F'=2$ transitions of 
the $^{87}$Rb $D1$ line. The 30~mW output of the laser is amplified by a 
tapered amplifier (TA) to about 1.8~W. Pulsed diffraction output from an 
acoustic-optical modulator (AOM) is coupled into a fiber-based electro-optical modulator (fEOM) for wide-band 
microwave modulation driven by a high-speed arbitrary waveform generator 
(Keysight M8195A), referred to as microwave AWG in the following. With AOM 
diffraction kept at a low duty cycle, the average laser power coupled into the fEOM is kept below 20~mW to avoid photo-refractive 
damage~\cite{Xuejian2018}.

To shape the optical pulses with microwave pulses, we use the side-band 
modulation technique by encoding the pulse shape information into amplitude 
$A(t)\in [0, 1]$ and phase $\varphi(t)$ of carrier modulation with 
$\omega_M$ angular frequency. The input-output relation for the complex 
electric field of the optical wave can be expressed as:
\begin{equation}
\begin{array}{l}
E_{\rm out}=e^{i \theta_0 A(t) \sin (\omega_M t +\varphi(t))}E_{\rm in} \\
~~~~~~=\sum_n i^n J_n(\theta_0 A(t)) e^{i n (\omega_M t+\varphi(t))} E_{\rm in}.
\end{array}\label{eqEOM}
\end{equation}
The phase modulation depth factor $\theta_0$ is decided by the microwave 
power and fEOM modulation efficiency.

\begin{figure*} \includegraphics[width=.8\textwidth]{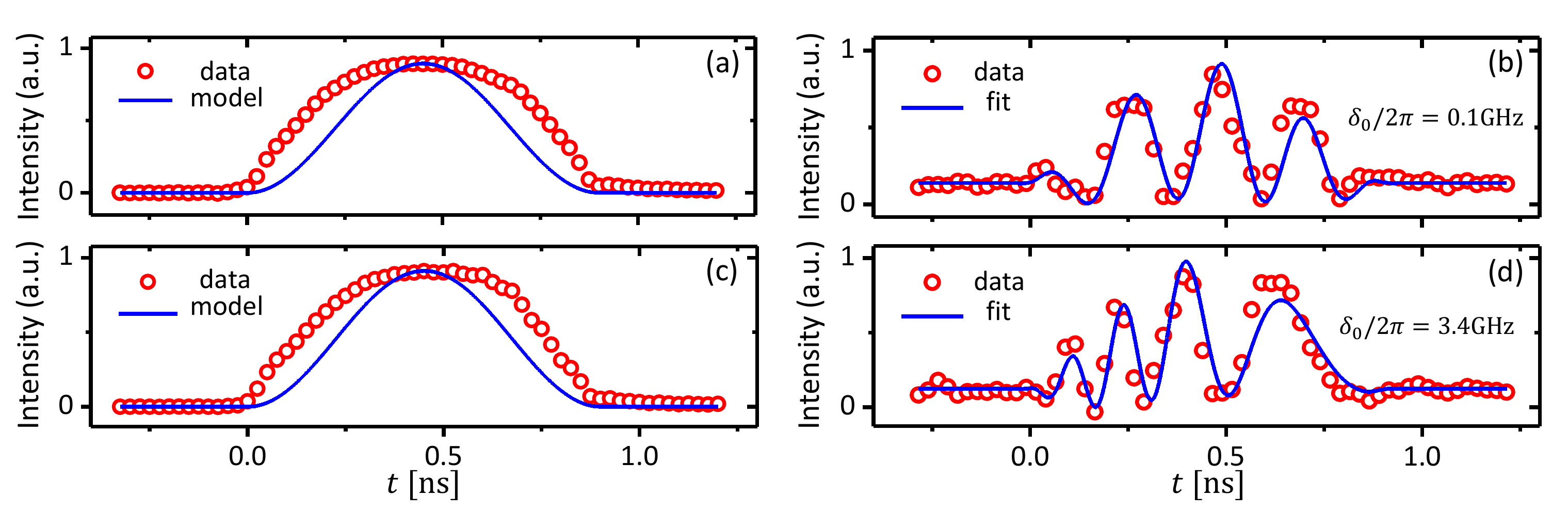} 
\caption{Intensity and beat note measurements of chirped pulses programmed 
with $s(t)\propto \sin^2(\pi t/\tau_c)$ and $\delta(t)=-\delta_0\cos(\pi 
t/\tau_c)$. (a) Intensity measurement (red points) and $\sin^2$-model (blue 
line) of resonant pulse with $\tau_c=0.9$~ns. The model is for the desired 
3$^{\rm rd}$ sideband of the fEOM output. The difference between the measurement 
and the model is mainly due to the $n=2,4$ sidebands detected by the 
photo-diode. (b) Beat signal (red points) and chirped-sine model fit (blue 
line) of the same pulse as in (a). The reference cw laser is 4.6~GHz blue 
detuned. The pulse is programmed with $\delta_0=0$~GHz, but the fit suggests 
$\delta_0=2\pi\times 0.1$~GHz, likely developed during the dispersive 
propagation of the pulse within the optical AWG. Similar data and analysis 
are presented in (c) and (d) for a chirped pulse with duration 
$\tau_c=0.9$~ns, with $\delta_0$ programmed to be $2\pi\times 4$~GHz but 
$\delta_0=2\pi\times 3.4$~GHz according to the fit. The intensity 
measurements in (a) and (c) are averaged for 4000 times, while the beat note 
signals in (b) and (d) are single-shot measurements.} 
\label{chirpedpulsesmeasurements}
\end{figure*}

The second line of Eq.~(\ref{eqEOM}) suggests we can simultaneously shape the 
amplitude and phase of $n^{\rm th}$-sideband with the single fEOM modulation. We 
choose $n=3$, $\omega_M=2\pi\times 16$~GHz, and adjust the microwave power 
toward $\theta_0\approx 4$ so as to maximize the modulation efficiency for 
the sideband decided by the Bessel function $J_3(\theta_0)$. To suppress the 
unwanted sidebands, the collimated fEOM output is sent through a $\sim 13$~GHz 
bandwidth optical filter, which is composed of a diffraction grating (2400 
lines/mm) and a single mode fiber. With the modulation efficiency limited by 
$|J_3(\theta_0)|^2 \sim 0.18$ and after all the coupling losses, we achieve 
20~mW peak power for the $n=3$ order sideband with modulation bandwidth 
limited by the grating filter. Due to the large $\omega_M$, the $n=2,4$ 
sidebands are less than $30\%$ of the total energy for typical pulse shapes 
[Figs.~\ref{chirpedpulsesmeasurements}(a)(c)]. The on-off power ratio is 
about $400:1$. The whole system is referred to as our $D1$ optical AWG.

In this work the $D1$ optical AWG serves to generate the sub-nanosecond 
chirped-sine pulses as in Fig.~1 in the main text but
 the actual pulse shape is expected to deviate due to the limited optical and 
microwave bandwidth and nonlinearity of the whole modulation system. 
Therefore, instead of assuming programmed pulse shapes, we directly measure 
the optical waveform to confirm its functional form, by beating the pulsed 
output with a reference cw laser which is $\Delta_0=2\pi\times 4.6$~GHz 
blue-detuned from the $F=2-F'=1,2$ crossover of the $^{87}$Rb $D1$ line. 
Typical intensity and beat note measurements are shown in 
Fig.~\ref{chirpedpulsesmeasurements}. The intensity measurements in 
Figs.~\ref{chirpedpulsesmeasurements}(a)(c) deviate from the $\sin^2(\pi 
t/\tau_c)$ model programmed for the 3$^{\rm rd}$ sideband,
mainly because of the unwanted sidebands (mainly the $n$=2 and $n$=4 orders) not 
fully suppressed by the grating filter.
However, we expect that the unwanted sidebands affect negligibly the $D1$ control 
due to their $\sim16$~GHz or larger detunings from the atomic resonance. The 
beat note signals as in Figs.~\ref{chirpedpulsesmeasurements}(b)(d) are 
quite well fitted by the interference expected from the chirped-sine form as 
$\sin(\pi t/\tau_c)\sin(\Delta_0 t-\delta_0 \tau_c \sin(\pi 
t/\tau_c)/\pi+\varphi_0)$. The fitted $\delta_0$ in this work are generally smaller than their programmed 
values. In addition to the $\delta_0$ calibration, we also calibrate the 
linearity of $\sqrt{s}$ ($s\equiv$ intensity$/I_{s1}$ is the saturation parameter and $I_{s1}= 4.49~$mW/cm$^2$ is the $D1$ transition saturation intensity) with respect to the programmed values according to 
the beat note measurements. An overall absolute value correction factor 
$\kappa$ multiplied to $\sqrt{s}$ is instead estimated by comparing 
simulation with experimental measurements of optical acceleration~\cite{coSub}.

Additional pulse distortion could come from possible sub-pulses due to 
multiple reflections at interfaces for the microwave and optical pulse 
propagation. To suppress electronically generated sub-pulses, care is taken 
to choose microwave cables with minimal lengths in this work. To suppress 
optically generated sub-pulses, the optical elements following the fEOM 
output are chosen to minimize unwanted retro-reflections. The absence of 
unwanted sub-pulses in the ${\bf k}_c$ beam is verified for delay beyond 
1~ns and for relative intensity beyond -40~dB,  with a multi-mode fiber 
coupled photon counter at nanosecond absolute time resolution. Within one 
nanosecond delay, any sub-pulse would lead to distortion of the nanosecond 
pulse shape itself. However, it is difficult to tell whether the small 
distortion as in Fig.~\ref{chirpedpulsesmeasurements} is indeed due to pulse 
distortion, or due to finite response of our fast photo detector (Thorlabs 
PDA8GS). By combining all the pulse measurements with an overall setup 
analysis,  we conclude that any sub-pulse co-propagating with the main pulse 
in our system is below 35~dB in relative intensity.

To increase the optical bandwidth of the fEOM-based laser system as in this work to, e.g.,~40~GHz level, one may accordingly increase the optical filter bandwidth, the fEOM carrier frequency, and its $n^{\rm th}$-order enhanced modulation bandwidth. A larger Rabi frequency can instead be achieved by more powerful pulsed seeding (at an even lower duty cycle), combined with better focusing of the laser to control a smaller sample. Instead of modulating a cw laser, a better approach to generate the wide-band optical waveforms with high peak power is likely based on precise shaping of mode-locked lasers~\cite{Yudi2020}.

The nanosecond $D2$ probe pulse is generated by another ECDL-fEOM setup that 
shares the same microwave AWG. The $D2$ optical AWG system also serves to 
generate the cooling laser~\cite{foot:YuzhuoLaser2019}. To ensure plenty of 
output power, the $D2$ laser is modulated by the fEOM before being amplified. 
This reversion of setup order introduces extra nonlinearity by the TA, 
leading to imperfect sine form of the probe pulse that is accounted for in our 
numerical modeling.

\subsection{Experimental Sequence\label{SecSeq}}

To produce the atomic sample in this work, up to $\sim10^7$ atoms are loaded 
into a magneto-optical trap (MOT) in less than 1~sec. Assisted by 
polarization gradient cooling, up to $10^5$ atoms are then loaded into a 
1064~nm crossed optical dipole trap at $\sim 0.5$~mK trap depth, which are 
subsequently transferred to a 840~nm dimple trap with up to $2\times 10^4$ 
atoms. This system is designed for evaporation of the sample to quantum 
degeneracy~\cite{DimpleDalibard2011}. In this work, slight evaporation in 
the hybrid trap produces the $\sim20~\mu$K atomic sample with up to $4\times 10^{12}$ cm$^{-3}$ peak density. The aspect ratio of the atomic sample is estimated with both imaging along the ${\bf e}_x$ and ${\bf e}_z$ direction (see the axes defined in Fig.~1 in the main text). In particular, the main measurement results are with approximate Gaussian $1/e$ radius of $(1.1,1.0,0.9) \sigma$ with $\sigma \approx 7~\mu$m along the $x,y,z$ axis respectively.

In the experiment, up to $N_{\rm rep}=$100 cycles of the probe-control sequence are applied to 
the atomic sample right after its release from the optical trap. Within each 
cycle, the probe pulse with central frequency resonant to $5S_{1/2} F=2 - 
5P_{3/2} F'=3$, and then the sequence of the $D1$ control pulses, are applied 
to the atomic sample. Synchronized with the probe pulse is an electronic 
trigger that starts a photon-counter to record the redirected fluorescence 
photons. We adjust the probe excitation strength $\theta_p$ so that typical 
counting probability per cycle $p < 0.25$ is small enough to avoid 
counter saturation. After each probe-control-measure cycle and a 
$\sim$100~ns delay, a 100~ns repumping pulse resonant to $F=1 - F'=2$ is 
applied to repump atoms in $F=1$ to $F=2$, with efficiency estimated to be 
better than $85\%$, before the next cycle.

The repeated superradiance measurements to the same atomic sample are 
accompanied by heating and loss of atoms that contribute to reduced 
collective emission signals. By comparing the intensity of superradiance for 
different repetition $N_{\rm rep}$, we found the superradiant intensity 
decreases to $\sim\{30\%,80\%,95\%\}$ with $N_{\rm rep}=\{100,50,10\}$ 
respectively. To find a balance between signal-to-noise and the atom-loss 
errors, for most of experiments in this work we only extract and average the 
signals from the $N_{\rm rep}=50$ repetitions of measurements. By further 
repeating the $N_{\rm rep}$-cycles $N_{\rm e}$ cold atom loading times, the 
overall measurement repetition is given by $N_{\rm exp}=N_{\rm e} N_{\rm 
rep}$, which is typically between 50000 and 70000 in this work.

\subsection{Hyperfine interaction during dipole spin-wave control}

We choose the polarization for the probe and control lasers to be along  ${\bf e}_y$ and ${\bf e}_x$ respectively. Taking the ${\bf e}_x-$direction as the quantization axis, the $\pi-$ control couplings to $5P_{1/2}$ would be with equal strengths and detunings for all the five 5$S_{1/2} F=2, m_F$ Zeeman sub-levels, and with vanishing hyperfine Raman coupling, if the 5$P_{1/2}$ hyperfine splitting $\Delta_{\rm D1, hfse}=2\pi\times 814.5$~MHz can be ignored. The approximation helps us to establish the simple two-level control picture in Fig.~1(d) in the main text even for the real atom.

Practically,  the hyperfine dephasing effects can be suppressed for atoms at $|m_F|=0,2$ Zeeman sublevels by adjusting the optical delay $\tau_d$  to match $2\pi/\Delta_{\rm D1, hfse}\approx 1.23~$ns. The hyperfine effect impacts more severely the $|m_F|=1$ atoms through both intensity-dependent dephasing and non-adiabatic population losses. These hyperfine effects are suppressible with better Zeeman state preparations, or by faster controls with $\tau_{c}\Delta_{\rm D1, hfse}\ll1$ while setting $\tau_d$ as multiples of $2\pi/\Delta_{\rm D1, hfse}$.

\section{Theoretical model\label{AppB}}
A full theoretical description of the light-atom interaction in this work 
involves dynamics of the fairly densely packed  multi-level atoms under the $D2$ weak excitation and $D1$ quasi-adiabatic control, with long-range 
interaction mediated by resonant exchange of photons. The full solution of 
such a multi-level, many-atom system remains an open theoretical and 
numerical challenge within quantum optics, which goes beyond the scope of 
this work. 
Here we focus on the spin-wave control itself that happens within nanoseconds with negligible interaction effects for our atomic sample with moderate density and atom number, so that a simple single-body picture becomes valid. Minimal modeling to account for cooperative emission can be found in ref.~\cite{coSub} which also include additional matterwave acceleration measurements and more detailed modeling of experimental parameters. We leave a detailed study of the (mis)matched excitation dynamics for a future work~\cite{shortPaper2}.


In the following we first briefly discuss the quasi-adiabatic control technique with a two-level model, including geometric and dynamic phases, and refer readers to refs.~\cite{Metcalf2005, nonAdiabaticMetcalf2007,AAgateZhuWangPRL2002,Gritsev2012} on more details of the control technique and its geometric aspects. Then we discuss the single-body simulation of the spin-wave dynamics which includes all the hyperfine levels of $D1$ and $D2$ transitions, so as to quantify experimental observations due to the nanosecond spin-wave control.

\subsection{Geometric phase patterning with optical rapid adiabatic passages\label{SecExpModel}}

We start with the interaction-free $N-$atom wavefunction to be expressed as $|\psi(t)\rangle=|\psi_1(t),\cdots,\psi_N(t)\rangle$, with $|\psi_i(t)\rangle$ describing single-body wavefunction for atom at location ${\bf r}_i$. We consider the auxiliary transition for the atom to be governed by a single-atom Hamiltonian as
\begin{equation}
    H_a^i=-\hbar\Delta|a_i\rangle\langle a_i|+ \frac{\hbar}{2}\big(\eta({\bf 
    r}_i)\Omega_{c}(t) e^{-i\varphi_{c}({\bf r}_i, t)}\sigma^+_{c,i}+h.c\big).
     \label{Ha1}
\end{equation}
Here we have $\sigma^+_{c,i}=|a_i\rangle\langle g_i|$. The 
time-dependent Rabi frequency $\eta({\bf r}_i) \Omega_{c}(t)=|{\bf E}_c({\bf 
r}_i,t)\cdot {\bf d}_{a g}|/\hbar$ is driven by the control laser with a 
Gaussian beam intensity profile, with $\Omega_c(t)$ the peak value and 
$\eta({\bf r}_i)\leq1$ a position dependent factor. The control phase 
$\varphi_c({\bf r},t)$ is written in the rotating frame detuned from the 
$|g\rangle-|a\rangle$ transition by $\Delta$. The goal is to design 
$\Omega_c(t)$, $\varphi_c({\bf r}_i,t)$ to achieve error-resilient state-dependent phase-patterning, ideally described by the single-body unitary operator 
\begin{equation}
U^i_g(\varphi_G({\bf 
r}_i))=|e_i\rangle\langle e_i|+ e^{i \varphi_G({\bf 
r}_i)}|g_i\rangle\langle g_i|
\end{equation}
for all the $N$ atoms within the $\{|g\rangle,|e\rangle\}$ subspace. Then the corresponding operation for the ensemble can be directly expressed as $U_g(\varphi_G) =\prod_{i=1}^N{U^i_g(\varphi_G({\bf r}_i))}$.

To implement the geometric phase patterning, we consider multiple pulse 
control as in this work with $\Omega_c(t)$ and $\varphi_c({\bf r}_i,t)$ 
split into $n$ smooth sections arriving at $t=t_n$. For a single pulse $n$, 
we consider $\dot \varphi_{c,n}({\bf r}_i,t)=\delta_{c,n}(t-t_n)$. By 
redefining $|a_i\rangle$ with $e^{-i\int^t \sum_n \delta_{c,n}(t'-t_n) d 
t'}$ phase factor,  Eq.~(\ref{Ha1}) can be rewritten in the new rotating 
frame as
\begin{equation}
\begin{array}{l}
H_a^i=-\frac{\hbar}{2} \sum_n \big(\Delta+\delta_{c,n}(t-t_n)\big)(\hat 
1_{g,a}+\sigma_{c,i}^z)+\\
~~~~~~~~~\big(\eta_n({\bf r}_i)\Omega_{c,n}(t-t_n) e^{-i\varphi_{n}({\bf 
r}_i)}\sigma^+_{c,i}+h.c\big),
 \end{array} \label{Ha2}
\end{equation}
with $\hat 1_{g,a}=|a_i\rangle\langle a_i|+|g_i\rangle\langle g_i|$ and 
$\sigma^z_{c,i}=|a_i\rangle\langle a_i|-|g_i\rangle\langle g_i|$.

Exploring the $\rm{SU}(2)$ symmetry in Eq.~(\ref{Ha2}), it is straightforward to 
achieve $U^i_g(\varphi_G)$, in absence of dipole-dipole interaction or 
spontaneous emission, by successively applying two nearly identical 
$|g\rangle-|a\rangle$ inversion pulses with $\delta_{c,1}=\delta_{c,2}$, 
$\Omega_{c,1}=\Omega_{c,2}$ but with different optical phase 
$\varphi_{1,2}({\bf r})$. In particular, we consider the state evolution 
$|\psi_i(t)\rangle$ with $|\psi_i(0)\rangle=|g_i\rangle$ subjected to 
$n=1,2$ population inversion pulses, each with a $\tau_c$ duration, and with 
$t_1=0$ and $t_2=\tau_d$. The ``return amplitude'' of cyclic evolution 
$f_g=\langle g_i|\psi_i(\tau_c+\tau_d)\rangle=|f_g|e^{-i\varphi({\bf r}_i)}$ 
is characterized by $\varphi({\bf r}_i)=\varphi_D({\bf r}_i)+\varphi_G({\bf r}_i)$ including 
dynamic  $\varphi_D$ and geometric $\varphi_G$ phases. For an ideal pair of 
population inversion pulses, $|f_g|=1$, and the $\rm{SU}(2)$ symmetry suggests
\begin{equation}
\varphi_G({\bf r}_i)=\pi+\varphi_1({\bf r}_i)-\varphi_2({\bf 
r}_i)\label{equiG}
\end{equation}
determined by the optical phase difference between the otherwise nearly 
identical pulse pair. The $\varphi_G$ is visualized on the Bloch sphere 
(Fig.~1 in main text) as half the solid angle spanned by the cyclic state 
trajectory. With the two-level symmetry, the dynamic phase 
$\varphi_D=\big(\int_0^{\tau_d}dt+\int_{\tau_d}^{\tau_d+\tau_c}dt\big)\langle 
\psi_i(t)|H_a^i|\psi_i(t)\rangle$ for the perfect inversions can be 
expressed as:
\begin{equation}
\begin{array}{l}
\varphi_D({\bf r}_i)= - \int_0^{\tau_d}\big(\Delta+\delta_c(t)\big) d t + \\
~~~~~~~~~~~~~\int_0^{\tau_c}\langle \psi_i(t)|(h_1-h_2)|\psi_i(t)\rangle d 
t,
    \end{array}\label{equiD}
\end{equation}
with $h_{1,2}$ given by the $n=1,2$ terms in the summation of 
Eq.~(\ref{Ha2}) excluding the $\hat 1_{g,a}$ part respectively.

To arrive at both Eqs.~(\ref{equiG}) and (\ref{equiD}), we assume $\eta_1\approx 
\eta_2$ and $|\psi_i(\tau_d<t<2\tau_d)\rangle$  approximately follows 
$|\psi_i(0<t<\tau_d)\rangle$ on the Bloch sphere up to a rotation. Any 
spatial-dependent $\varphi_{D,a}({\bf r}_i)$ is nullified if the two 
inversion pulses are with identical strength so that $\eta_1=\eta_2$. The 
additional, spatially independent phase 
$\varphi_{D,d}= - \int_0^{\tau_d}\big(\Delta+\delta_c(t)\big) d t$ is usually 
harmless for ensemble control of two-level atoms. Here for the 
$|g\rangle-|e\rangle$ dipole control, however, $\varphi_{D,s}$ needs to be 
tuned to $2\pi$-multiples, particularly if multiple choices of $\Delta$ 
exist for the $|g\rangle-|a\rangle$ transition, such as those due to 
hyperfine splitting in this work.

To achieve $\Delta$ and $\eta({\bf r}_i)$ independent population inversion, 
the simplest choice is a quasi-adiabatic pulse. With $\Omega_c=\Omega_0 
\sin(\pi t/\tau_c)$ and $\delta_c=-\delta_0 \cos(\pi t/\tau_c)$, stability 
of near unity inversion efficiency against $\Delta$ and $\eta$ has been 
studied in detail in the context of nuclear magnetic 
resonance~\cite{NMRAdiabatic2001}, molecular spectroscopy~\cite{Loy1974}, 
and matter-wave accelerations~\cite{Metcalf2005,nonAdiabaticMetcalf2007}. 
Efficient and error-resilient inversion is achievable with 
$(\Omega_0,\delta_0)$ close in magnitude and for $\int \Omega_c d t$ beyond 
$3\pi$, as in this work.

In addition, we model the $D2$ spin-wave excitation with single atom Hamiltonian
\begin{equation}
    H_e^i=-\hbar\Delta_e |e_i\rangle\langle e_i|+ \frac{\hbar}{2}\big(\Omega_{p}(t) e^{-i\varphi_{p}({\bf r}_i)} \sigma^+_{i} +h.c\big)
     \label{He1}
\end{equation}
where $\Omega_{p}$ is the Rabi frequency of the probe pulse with $\int_{-\tau_p}^{0} \Omega_p d t \ll 1$ and $\sigma^+_{i}=|e_i\rangle\langle g_i|$. $\varphi_{p}({\bf r}_i) = -{\bf k}_p \cdot {\bf r}_i$ is the plane-wave optical phase of the probe pulse.  

\subsection{Perturbative estimation of interaction induced control error}
The resonant dipole interaction among atoms can be described by the following interaction potential~\cite{selectiveRadiancePRXchang17,coSub}:

\begin{equation}
\begin{array}{l}
    \hat V^{i,j}_{\rm D D}=-\frac{k_p^2}{\varepsilon_0} {\bf d}^*_{e g} \cdot {\bf G}({\bf r}_i-{\bf r}_j,\omega_{eg}) \cdot {\bf d}_{e g} \sigma^+_i \sigma^-_j+\\
    ~~~~~~~~~-\frac{k_c^2}{\varepsilon_0} {\bf d}^*_{a g} \cdot {\bf G}({\bf r}_i-{\bf r}_j,\omega_{ag}) \cdot {\bf d}_{a g} \sigma^+_{c,i} \sigma^-_{c,j}, 
    \end{array} \label{eqRDI2}
\end{equation}
with $\sigma^-_{i}=(\sigma^+_{i})^{\dagger}$ and $\sigma^-_{c,i}=(\sigma^+_{c,i})^{\dagger}$. Here $\omega_{eg}$, $\omega_{ag}$ are optical transition frequencies and ${\bf d}_{e g}$, ${\bf d}_{a g}$ are transition matrix elements of the $|g\rangle-|e\rangle$ and $|g\rangle-|a\rangle$ transitions, respectively. And correspondingly we have $k_p = \omega_{eg} / c$ and $k_c = \omega_{ag} / c$. $\varepsilon_0$ is the vacuum permittivity. We refer the first and second line as $V^{i j}_{{\rm DD},eg}$ and $V^{i j}_{{\rm DD},ag}$ respectively. ${\bf G}({\bf r},\omega)$ is the free-space Green's tensor of the electric field obeying 
\begin{equation}
\nabla\times\nabla\times {\bf G}({\bf r},\omega)-\frac{\omega^2}{c^2} 
{\bf G}({\bf r},\omega)=\delta^3({\bf r}) \mathds{1}.
\end{equation}

We perturbatively estimate the influence of resonant-dipole interaction by Eq.~(\ref{eqRDI2}) to the geometric phase patterning of collective dipole excitation. For simplicity we only consider the ``timed-Dicke'' state $|\psi(0)\rangle=|\psi_{\rm TD}({\bf k}_p)\rangle$ (as defined in the main text). With $\varphi_G({\bf r}_i)=\pi+2{\bf k}_c\cdot {\bf r}_i$, a perfect phase-patterning operation leads to $|\psi(\tau_c+\tau_d)\rangle=U_g(\varphi_G)|\psi_{\rm TD}({\bf k}_p)\rangle=e^{i \phi_0}|\psi_{\rm TD}({\bf k}_p-2{\bf k}_c)\rangle$ in absence of $V^{i,j}_{{\rm DD}}$, where $\phi_0 =  (N-1)\pi + \sum_{i=1}^N{2 {\bf k}_c \cdot {\bf r}_i} $ is a global phase. Error associated with the resonant dipole interaction can be estimated with either incoherent phase error $\delta \varphi_{\rm inc}=\sqrt{\overline{\delta \varphi^2}}$, or the collective phase error $\delta \varphi_N=\sqrt{N\overline{\delta \varphi^2}}$, with 
\begin{equation}
\begin{array}{l}
    \overline{\delta \varphi_e^2} = \sum_{i}\big|\sum_{j}\int_0^{\tau_c+\tau_d} \langle \psi(t)|V^{i j}_{{\rm DD},eg}|\psi(t)\rangle d t\big|^2,\\
\overline{\delta \varphi_a^2} =\sum_i\big|\sum_{j}\int_0^{\tau_c+\tau_d} \langle  \psi(t)|V^{i j}_{{\rm DD},ag}|\psi(t)\rangle d t\big|^2
\end{array}\label{equError1}
\end{equation}
for contribution from the $V^{i j}_{{\rm DD},eg}$ and $V^{i j}_{{\rm DD},ag}$ interaction respectively. The error is evaluated with the unperturbed $|\psi(t)\rangle$ evolving according to the  single-body $H_{c1},H_{c2}$ control. 
The perturbative treatment is valid for $\delta \varphi\ll1$, which is generally achievable with increased $(\Omega_0,\delta_0)$ and reduced $\tau_d+\tau_c$ control interval. 

We consider the ``worst case scenario'' where control errors due to all the pairwise interactions add up coherently to perturb the collective control dynamics, with overall error characterized by $\delta \varphi_N$. We also consider the shortest possible duration $\tau_c+\tau_d=2\tau_c$. In light of the fact that the collective error is associated with collective interaction, it is straightforward to have $\delta \varphi_{N,e}\sim \tau_c \sqrt{\langle \delta_{N,eg}\rangle^2+\langle\Gamma_{N,eg}\rangle^2/4}$, that scales with the largest collective Lamb shift $\delta_{N,eg}$ and decay rate $\Gamma_{N,eg}$ of the singly excited gas~\cite{clScullyPRL09,Kaiser2016,coSub}. With the $|g\rangle -|a\rangle$ population inversions and phase imprinting, $\delta_{N,eg}$ and $\Gamma_{N,eg}$ reduce substantially during the control. The symbol $\langle ...\rangle$ averages over the instantaneous values, leading to at least a factor of $50\%$ reduction to the collective part of $\delta \varphi_{N,e}$. 


Resonant dipole interaction on the strongly driven $|g\rangle-|a\rangle$ transition is much stronger than on the weakly excited $|g\rangle-|e\rangle$ transition. Accordingly, $\delta \varphi_{N,a}\sim N \tau_c \sqrt{\langle \delta_{N,ag}\rangle^2+\langle\Gamma_{N,ag}\rangle^2/4}$ can be much larger. However, in contrast to $\delta \varphi_{N,e}$, it is not appropriate to simply associate $\delta \varphi_{N,a}$ with control error since the collective radiation addresses the same $|g\rangle-|a\rangle$ transition as the ``very strong'' control $\Omega_c$ does. In fact, during the control the collective dipole radiation amounts to absorbing and reshaping the $\Omega_c({\bf r}, t)e^{-i\varphi({\bf r}, t)}$ control pulse, and the adverse effects can be largely compensated for by quasi-adiabatic techniques insensitive to the pulse shape for population inversions~\cite{foot:odPi}. The collective radiation thus impacts the phase patterning operation as a collective dynamic phase shift according to Eq.~(\ref{equiD}), which can be quite uniform across the atomic sample and do not contribute to the actual collective dipole control error. With a concrete study of the open system coherent control for future work~\cite{Koch2016}, we conclude this section by suggesting that the control error due to resonant dipole interactions depends on details of atomic position arrangements, and could be bounded by $\delta \varphi_M \sim {\max}(\delta \varphi_{N,e}, \delta \varphi_{i,a})$ with careful choice of $\Omega_c,~\varphi_G$ to avoid substantially distortion of the shape-optimized control pulses by the atoms, or simply by implementing the control instead on excited state transitions~\cite{multiphotonMOT} in which case $\delta \varphi_a$ becomes much less important.

\subsection{Simulation of spin-wave dynamics supported by non-interacting atoms}

\begin{figure} \includegraphics[width=0.43\textwidth]{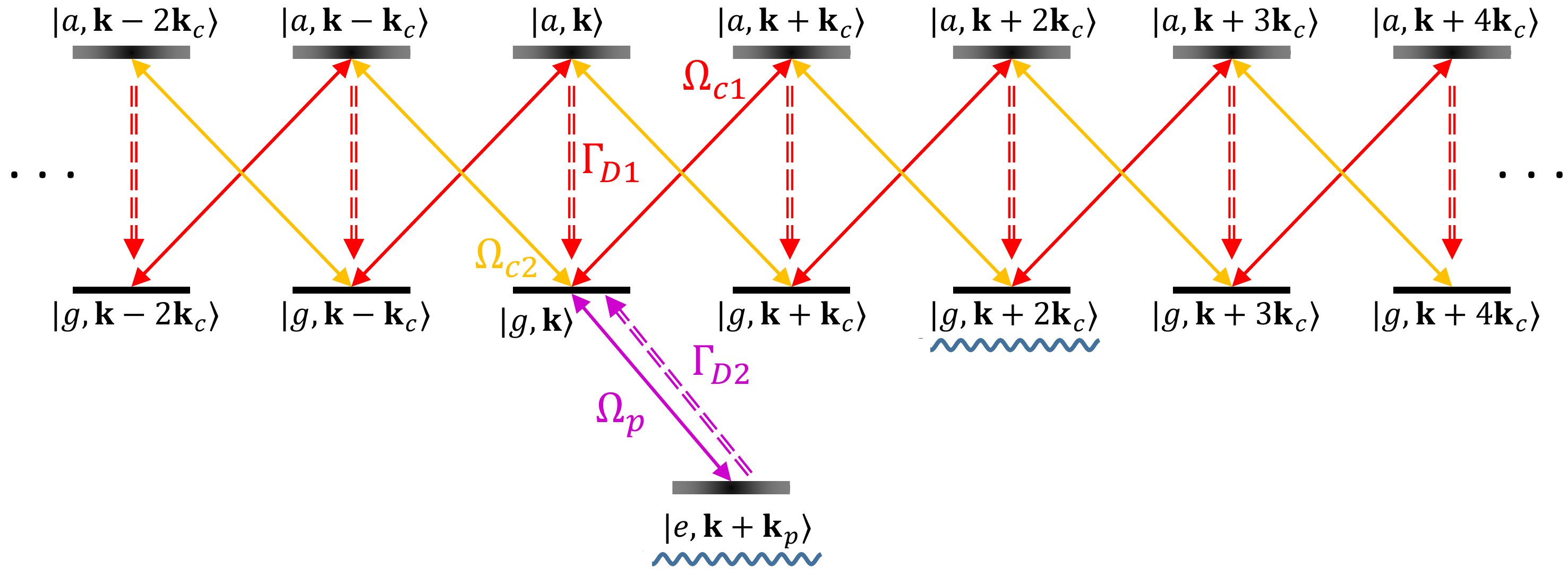} 
\caption{``Momentum lattice'' structure for probe excitation and $U_g$ 
control simulations according to 
Eqs.~(\ref{equHeff3}), (\ref{equiC}), and (\ref{equiMasterS}). Dashed arrows represent 
the ``effective'' quantum jump operations associated with Eq.~(\ref{equiC}). 
The double-sided arrows represent the coherent laser couplings. The 
coherence between the wavy
 underlined lattice sites $|e,{\bf k}+{\bf k}_p\rangle$ and $|g,{\bf 
k}+2{\bf k}_c\rangle$ is associated with the redirected superradiant 
emission.} \label{figMLattice}
\end{figure}

We now ignore resonant dipole interactions among atoms, and use single atom optical Bloch equations to simulate the spin wave dynamics for the experiments. It is important to point out that the simple 
single-body method is incapable of describing initially entangled states 
such as timed-Dicke states. However, as long as the 
observables to be evaluated are only composed of linear dipole operators and the interaction effects are negligible, the same dynamics should be captured by the product states of weakly excited single atoms in the linear optics regime. 


We write down the effective non-Hermitian Hamiltonian for the interaction-free model
\begin{equation}
    H'_{\rm eff}=\sum_{j=1}^N 
\big(H^j_a+H^j_e-i\hbar\frac{\Gamma_{\rm D2}}{2}|e_j\rangle\langle 
e_j|-i\hbar\frac{\Gamma_{\rm D1}}{2}|a_j\rangle\langle a_j|\label{eqHeff2}\big).
\end{equation}
Here $H^j_a$ and $H^j_e$ are according to Eq.~(\ref{Ha1}) and Eq.~(\ref{He1}), respectively. With the simplified Hamiltonian in Eq.~(\ref{eqHeff2}) that ignores 
atom-atom interaction, we are now free to choose ${\bf r}_i$ for notation 
convenience, in particular, we change the basis for single atom wavefunction 
into ${\bf k}-$space, with $|g,{\bf k}\rangle=\frac{1}{\sqrt{N}}\sum_i e^{i{\bf 
k}\cdot {\bf r}_i}|g_i\rangle$ and similarly for $|e,{\bf k}\rangle$ and 
$|a,{\bf k}\rangle$, by adjusting ${\bf r}_i$ to ensure the orthogonality of 
the ${\bf k}-$state basis of interest. In addition, we expand the level structure 
to that of the $^{87}$Rb $D1$ and $D2$ line, and use $\{g,e,a\}$ as indices to 
label the $\{5S_{1/2},5P_{3/2},5P_{1/2}\}$ hyperfine levels respectively. We 
end up with $H_{\rm eff}=\sum H^{(s)}_{\rm eff}$ composed of single-atom 
Hamiltonian in ${\bf k}-$space, with
\begin{equation}
    \begin{array}{l}
H^{(s)}_{\rm eff}=H_p+H_{c1}+H_{c2},\\
\\
H_p=\hbar\sum_{g,{\bf k}}\Delta_g |g,{\bf k}\rangle\langle g,{\bf k}|+\\
~~~~~~~\hbar\sum_{e,{\bf k}} (-\Delta_e-i\Gamma_{\rm D2}/2)|e,{\bf k}\rangle\langle 
e,{\bf k}|+\\
~~~~~~~\hbar\sum_{a,{\bf k}}(-\Delta_a-i\Gamma_{\rm D1}/2)|a,{\bf k}\rangle\langle 
a,{\bf k}|+\\
~~~~~~~\hbar\sum_{g,e,{\bf k}}\big(\frac{1}{2}\Omega_{p}(t) c^y_{e g} 
|e,{\bf k}+{\bf k}_p\rangle \langle g,{\bf k}|+h.c.\big),\\
\\
H_{c1} = - \hbar\sum_{a,{\bf k}}\delta_{c1}(t)|a,{\bf k}\rangle\langle a,{\bf k}|+\\
~~~~~~~~\hbar\sum_{g,a,{\bf k}}\big(\frac{1}{2}\eta_1 \Omega_{c1}(t) c^x_{a 
g} |a,{\bf k}+{\bf k}_c\rangle \langle g,{\bf k}|+h.c.\big),\\
\\
H_{c2} = - \hbar\sum_{a,{\bf k}}\delta_{c2}(t)|a,{\bf k}\rangle\langle a,{\bf k}|+\\
~~~~~~~~\hbar\sum_{g,a,{\bf k}}\big(\frac{1}{2}\eta_2 \Omega_{c2}(t) c^x_{a 
g} |a,{\bf k}-{\bf k}_c\rangle \langle g,{\bf k}|+h.c.\big).
\end{array}\label{equHeff3}
\end{equation}
Here, by properly setting the functions $\Omega_{c1}(t)$, $\Omega_{c2}(t)$, $\Omega_{p}(t)$, $\delta_{c1}(t)$ and $\delta_{c2}(t)$, we can encode the ``re-direction''--``switch-off''--``recall'' sequence in the Hamiltonian, which is shown in the Fig.~1(c) in the main text. The $c^x_{ag}$,$c^y_{eg}$ are 
combinations of  Clebsch-Gorden coefficients to characterize the $D1$ and $D2$ 
transitions driven by the ${\bf e}_x-$ and ${\bf e}_y-$ polarized control and 
probe lasers, respectively. We also introduce $\eta_{1,2}$ factors similar to 
those in Eq.~(\ref{Ha1}) to account for laser intensity inhomogenuities.

\begin{figure} 
\includegraphics[width=0.43\textwidth]{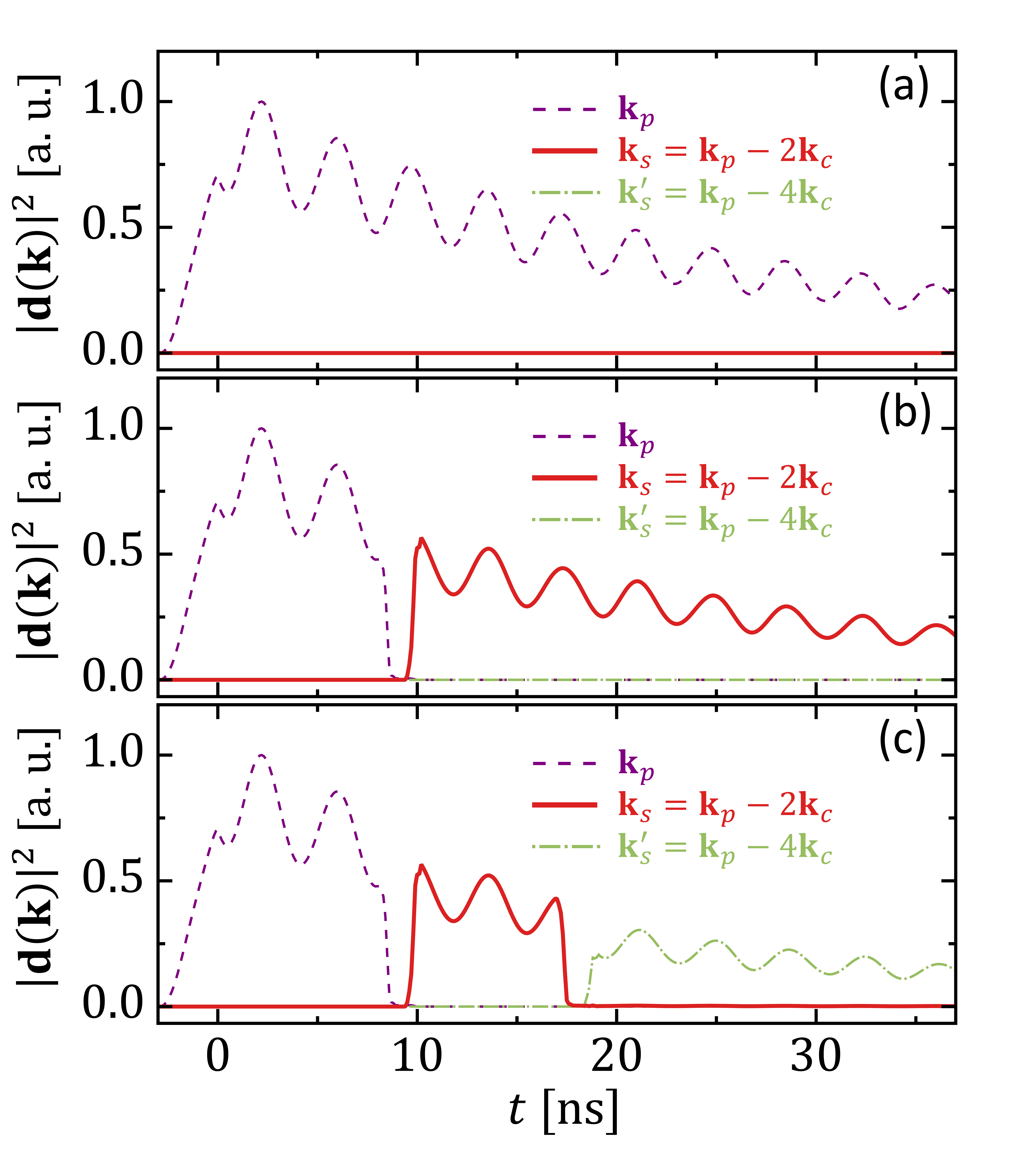} 
\caption{Simulation of collective dipole dynamics for typical spin-wave control sequences. The probe excitation is in the interval $-\tau_p<t<0$, with $\tau_p=3$~ns. (a) No $D1$ control is implemented. (b) The $D2$ probe excitation is followed by a $U_g(\varphi_G)$ control composed of two chirped $D1$ pulses with 
$\tau_c=0.9$~ns, $\tau_d=1.36$~ns, $\Omega_0 = 2\pi\times 3$~GHz and $\delta_0 = 2\pi\times 3.4$~GHz at $\Delta t_1=8$~ns. (c) A 
second $U_g$ operation at $\Delta t_2= 6.6 $~ns is implemented.  
The $|{\bf d}({\bf k}_p)|^2$, $|{\bf d}({\bf k}_s)|^2$ and $|{\bf d}({\bf k}_s')|^2$ are plotted with dashed, solid and dash-dotted lines, respectively. 
} \label{simulatefigureonoff}
\end{figure}

Clearly, Equation~(\ref{equHeff3}) can also be interpreted as being written 
in momentum space with quantized atomic wavefunctions, without kinetic energy 
terms. Indeed, the atomic motion within the sub-nanosecond control in this 
work can be ignored, and we adapt this wavefunction interpretation when 
using the same equations to calculate both dipole control and optical 
acceleration~\cite{coSub}.

To write down the single-atom master equation, we introduce six 
``effective'' collapse operators
\begin{equation}
\begin{array}{l}
    \hat C^j_{\rm D1}=\sum_{a,g,{\bf k}} \sqrt{\Gamma_{\rm D1}}c^{j}_{a g} 
|g,{\bf k}\rangle \langle a,{\bf k}|.\\
    \hat C^j_{\rm D2}=\sum_{e,g,{\bf k}} \sqrt{\Gamma_{\rm D2}}c^{j}_{e g} 
|g,{\bf k}\rangle \langle e,{\bf k}+{\bf k}_p|
\end{array}\label{equiC}
\end{equation}
with ``$j$'' running through ``$x$'',``$y$'' and ``$z$'' polarizations. The 
collapse operators are associated with quantum jumps and spontaneous 
emission. We effectively set the recoil ${\bf k}-$shifts in simple ways to 
minimize the calculation complexity, without affecting the $D_2$ dipole 
coherence and the $D_1$ optical force under study.

We are now able to write down the master equation for the single-atom 
density matrix $\rho^{(s)}$ as
\begin{equation}
\begin{array}{l}
\dot{\rho}^{(s)}(t)=\frac{1}{i} (H^{(s)}_{\rm eff} \rho^{(s)} -\rho^{(s)} 
H_{\rm eff}^{(s)\dagger} )+\\
~~~~~~~~~~~~~~~~~\sum_{j} (\hat{C}^j_{\rm D1} \rho^{(s)} 
\hat{C}_{\rm D1}^{j\dagger}+\hat{C}^j_{\rm D2} \hat{\rho} \hat{C}_{\rm D2}^{j\dagger}).
\end{array}\label{equiMasterS}
\end{equation}

With $\rho^{(s)}(t)$ it is straightforward to calculate the interaction-free 
evolution of the many-atom density matrix $\rho(t)=(\rho^{(s)}(t))^{\otimes N}$ 
and to evaluate collective observables $\langle \hat O \rangle={\rm 
tr}(\rho(t) \hat O)$.

We further simplify Eq.~(\ref{equHeff3}) by restricting the momentum basis 
according to situation of our experiments. The ${\bf k}$ states are coupled to ${\bf k}+ {\bf k}_p$ and ${\bf k}\pm n {\bf k}_c$ states via the probe and control interactions. By ignoring atomic motion, we only consider a single ${\bf k}-$class as illustrated by the ``momentum lattices'' in Fig.~\ref{figMLattice} which also highlight the structure of the couplings according to 
Eqs.~(\ref{equHeff3}), (\ref{equiC}), and (\ref{equiMasterS}). We restrict the 
accessible momentum states with $|n|<10$ for the numerical calculations.  The 
truncation is validated by numerically monitoring the high-$n$ states and by 
verifying the consistent results with larger $n-$cutoffs.

For simplicity and without loss of generality~\cite{coSub}, we set $\eta_1 = \eta_2 = 1$ and obtain the numerically evaluated single-atom density matrix $\rho^{(s)}(t)$ according to Eq.~(\ref{equiMasterS}). We further evaluate the dipole coherence ${\bf d}({\bf k}_s) = {\rm tr}\big(\rho^{(s)} (t){\bf d}^{-}({\bf k}_s)\big)$ for the weakly and coherently excited gas, with coherence operator ${\bf d}^{-}({\bf k}_s)={\bf e}_y \sum_{g,e}c^{y}_{e g}|g,{\bf k}+2 {\bf k}_c\rangle\langle e,{\bf k}+{\bf k}_p|$. Similarly, we can define ${\bf d}^{-}({\bf k}_p)={\bf e}_y \sum_{g,e}c^{y}_{e g}|g,{\bf k}\rangle\langle e,{\bf k}+{\bf k}_p|$ and ${\bf d}^{-}({\bf k}_s')={\bf e}_y \sum_{g,e}c^{y}_{e g}|g,{\bf k}+4 {\bf k}_c\rangle\langle e,{\bf k}+{\bf k}_p|$. The collective coherence in 
the main text is related to $\langle {\bf d}^{-}({\bf k}_s)\rangle$ as $\langle 
S^{-}({\bf k}_s) \rangle\propto N\langle {\bf d}^{-}({\bf k}_s)\rangle$, and 
furthermore we approximate $\langle S^{+}({\bf k}_s)S^{-}({\bf k}_s) \rangle 
\approx |\langle S^{-}({\bf k}_s)\rangle|^2$ for $N\gg1$. We then compare the numerically calculated $|{\bf d}({\bf k}_s)|^2$ ($\propto \langle S^{+}({\bf k}_s)S^{-}({\bf k}_s) \rangle$) with the experimentally observed superradiant signal $i_{{\bf k}_s}$.

A simulation of a full ``re-direction''--``switch-off''--``recall'' sequence is already given in Fig.~3a of the main text. As another illustrative example, here Fig.~\ref{simulatefigureonoff} presents the numerically evaluated $|{\bf d}({\bf k})|^2$ with ${\bf k} = {\bf k}_p$, ${\bf k}_s$ and ${\bf k}_s'$ for typical spin-wave control sequences including probe excitation, ``re-direction'' and ``switch-off''.  In all the simulation, we set the initial state to be $\rho^{(s)}_0 = \frac{1}{5}\sum_{g=1}^5 |g,{\bf k}\rangle\langle g,{\bf k}|$ with $|g\rangle$ running through the $|F=2,m_F\rangle$ Zeeman sublevels.  We notice the accuracy of the single-body simulation is reduced when simulating the spin-wave decays on the tens of nanoseconds time scales since the atom-atom interaction effects becomes important. Indeed the interactions lead to the cooperatively enhanced emission rate as observed experimentally and discussed in ref.~\cite{coSub}. 


\begin{thebibliography}{99}
\bibitem{foot:YuzhuoLaser2019}
The work on multi-frequency cooling laser system based on single EOM 
modulation and optical filtering will be for a future publication.
\bibitem{shortPaper2}
The spin wave control technique allows us to precisely characterize the phase-(mis)matched dipole spin waves dynamics as a function of atomic density and number. A paper to present the findings is under preparations.
\end{thebibliography}
\end{document}